\documentclass[12pt]{iopart}
\usepackage{graphicx}

\begin{document}

\title[Effect of disorder with long-range correlation on transport in graphene nanoribbon]
{Effect of disorder with long-range correlation on transport in graphene nanoribbon}
\author{G. P. Zhang$^{1,2}$, M. Gao$^{1}$, Y. Y. Zhang$^{3}$, N. Liu$^{4}$, Z. J. Qin$^{5}$, and M. H. Shangguan$^{1}$}
\address{$^{1}$Department of Physics, Renmin University of China, Beijing 100872, China}
\address{$^{2}$Ames Laboratory-US DOE, and Department of Physics and Astronomy, Iowa State University, Ames, Iowa 50011, USA}
\address{$^{3}$State Key Laboratory of Superlattices and Microstructures, Institute of Semiconductors, Chinese Academy of Sciences, P.O. Box 912, Beijing 100083, China}
\address{$^{4}$School of Microelectronics and Solid State Electronics, University of Electronic Science and Technology of China, Chengdu 610054, China}
\address{$^{5}$School of Physics and Engineering, Zhengzhou University, Zhengzhou 450001, China}
\ead{zhanggp96@ruc.edu.cn}
\date{\today}

\begin{abstract}
Transport in disordered armchair graphene nanoribbons (AGR) with long-range correlation between quantum wire contact is investigated by transfer matrix combined with Landauer's formula. Metal-insulator transition is induced by disorder in neutral AGR. Thereinto, the conductance is one conductance quantum for metallic phase and exponentially decays otherwise, when the length of AGR approaches to infinity and far longer than its width. Similar to the case of long-range disorder, the conductance of neutral AGR first increases and then decreases while the conductance of doped AGR monotonically decreases, as the disorder strength increases. In the presence of strong disorder, the conductivity depends monotonically and non-monotonically on the aspect ratio for heavily doped and slightly doped AGR respectively.
For edge disordered graphene nanoribbon, the conductance increases with the disorder strength of long-range correlated disordered while no delocalization exists, since the edge disorder induces localization.
(Some figures in this article are in colour only in electronic version)
\end{abstract}

\maketitle

\section{Introduction}
Graphene, discovered in 2004 \cite{a1,a2,a3,a4}, stimulated intensive research interest in fundamental properties and its potential application in nanoelectronic device due to its superior properties such as high mobility \cite{a3} and heat dissipation \cite{graphene-heat}. For an infinite pure graphene sheet, the electronic spectrum is linear around Dirac point, and can be described as massless Dirac fermions. Due to its unique electronic structure, many interesting phenomena had been observed such as half-integer quantum Hall effect \cite{a4} and bipolar current \cite{bipolar-current}. Now it is commonly thought that the most possible application may come from graphene nanoribbon, and the basic types of graphene nanoribbon are armchair-edged and zigzag-edged ones. Defects and impurities are inevitable in graphene-based materials, therefore effects of defects, impurities and disorder on transport properties are of application interest and had been reviewed in Refs. \cite{transport-disorderedgraphene-review,graphene-review,disorderedgraphene-review,mono-bilayer-review}. As for disorder, it is usually classified as short- and long-range disorder depending on the ratio of the interaction range to the lattice distance \cite{Rycerz-EPL2007}. Anderson disorder is a typical short-range disorder, and localization is induced by strong disorder. Long-range disorder such as Guassian-type disorder \cite{ZhangYY-PRL2009} induces metal-insulator transition at Dirac point. After several years of intensive investigation, it was well known that the scattering between two nonequivalent valleys is absent in the presence of long-range disorder \cite{disorderedgraphene-review}. Interestingly, off-diagonal disorder \cite{XiongSJ-PRB2007,QinZJ-JPCM2011} enhances electronic transport and superconductivity correlation at Dirac point. Recently, more and more attentions were paid to the effect of charged impurities on the transport in graphene based materials \cite{Hwang-chargedimpurity-RPL2007,Chen-chargedimpurity-NatPhys2008} since the observation of electron-hole puddles in graphene \cite{electron-hole-puddles}.

It was commonly believed that Anderson localization \cite{Anderson-PR1958} in one- and two-dimensional systems is induced by even a weak disorder from the scaling theory \cite{Anderson-PRL1979}. It was found that one or several resonant states may occur in one-dimensional disordered chain with short-range correlation \cite{Wu-shortPRL1991,Philips-shortScience1991,ChenXS-PLA1993}. Delocalized states within a continuous energy range exist in one-dimensional system, when disorder in onsite energy is long-range correlated \cite{FABF-longPRL1998,Izrailev-RPL1999}, where the site energies have an approximate spectral density of the form $s(k) \propto k^{-\alpha}$ with $s(k)$ being the Fourier transformation of the two-point correlation function of site energies $<\epsilon_{i}\epsilon_{j}>$ and $k$ the inverse of the wavelength $\lambda$ of the undulation on the site-energy landscape. The scaling behavior of the resistivity verified that delocalized states persist in the thermodynamic limit, and the relation between disorder correlation and the resistivity proved that disorder correlation is important for delocalization \cite{ZhangGP-EPJB2002}. There exists a Kosterlitz-Thouless type metal-insulator transition in two dimensional systems with long-range correlated disorder \cite{LiuWS-JPCM1999,LiuWS-EPJB2003}. Nowadays, the effect of disorder with long-range correlation on transport is under intensive investigation in many fields, which recently had been reviewed in Ref. \cite{Izrailev-PhysRep2011}.

Recently, the effect of disorder with long-range correlation in graphene attracts many researchers' attention \cite{LiQZ-PRL2011,Yan-RPL2011,Cheraghchi-PRB2011,DiracFermion-arxiv}. For example, transport in disordered graphene superlattice with long-range correlation was studied on basis of massless Dirac equation \cite{Cheraghchi-PRB2011}. It was found that there exists a truly metallic phase at the thermodynamic limit in the presence of long-range correlation, and the incident angle of metallic state increases with the correlation strength. Furthermore the metallic phase was obtained in term of correlation strength and disorder strength.

It was found that there exists equivalent mode selection between armchair graphene nanoribbon (AGR) and quantum wire (QW) \cite{AGR-QW-equivalent}. However, there is no such equivalence between zigzag graphene nanoribbon (ZGR) and QW, and even-odd parity occurs when ZGR is connected to QW contact even though ZGR is metallic \cite{ZhangGP-PLA2010,ZhangGP-CPL2011}. Moreover, the conductance of graphene nanoribbon at Dirac point is independent of the contact structure, provided that there are dense modes around Dirac point \cite{ZhangGP-CPL2011,densecontact}. Similar to our previous work \cite{ZhangGP-CPL2011,ZhangGP-PLA2011}, QW is adopted to mimic metallic contacts here. In this Letter, we investigate the transport of disordered AGR with long-range correlation based on tight binding model by transfer matrix method. Our main results are listed as follows. Localization-delocalization transition is induced by the long-range correlation of disorder when the disorder is weaker than one certain critical value. Thereinto, the conductance for metallic phase approaches one conductance quantum and exponentially decays otherwise, as the length increases to infinity and is far longer than the width. Similar to the case of long-range disorder \cite{Rycerz-EPL2007}, the conductance of neutral AGR first increases and then decreases while the conductance of doped AGR monotonically decreases, as the disorder strength increases. For disordered AGR with moderate aspect ratio at strong disorder, $\beta$ varies from negative to positive in the scaling function of conductivity $O(L^{\beta})$ outside and within a certain range of energy shift respectively. Therefore the conductivity depends monotonically and non-monotonically on the aspect ratio for heavily doped and slightly doped AGR respectively.

The structure of this Letter is as follows. In Sec. II, we introduce the structure of AGR between QW contacts, the tight binding Hamiltonian and long-range correlated disorder; in Sec. III, we show the dependence of the conductance on the strength of correlation and disorder, and the dependence of the conductivity on the aspect ratio in the presence of strong disorder that induces Anderson localization at the thermodynamic limit; and in Sec. IV, we make a summary.

\section{Structure and formalism}
\begin{figure}[tbh]
\includegraphics[width=8.0cm]{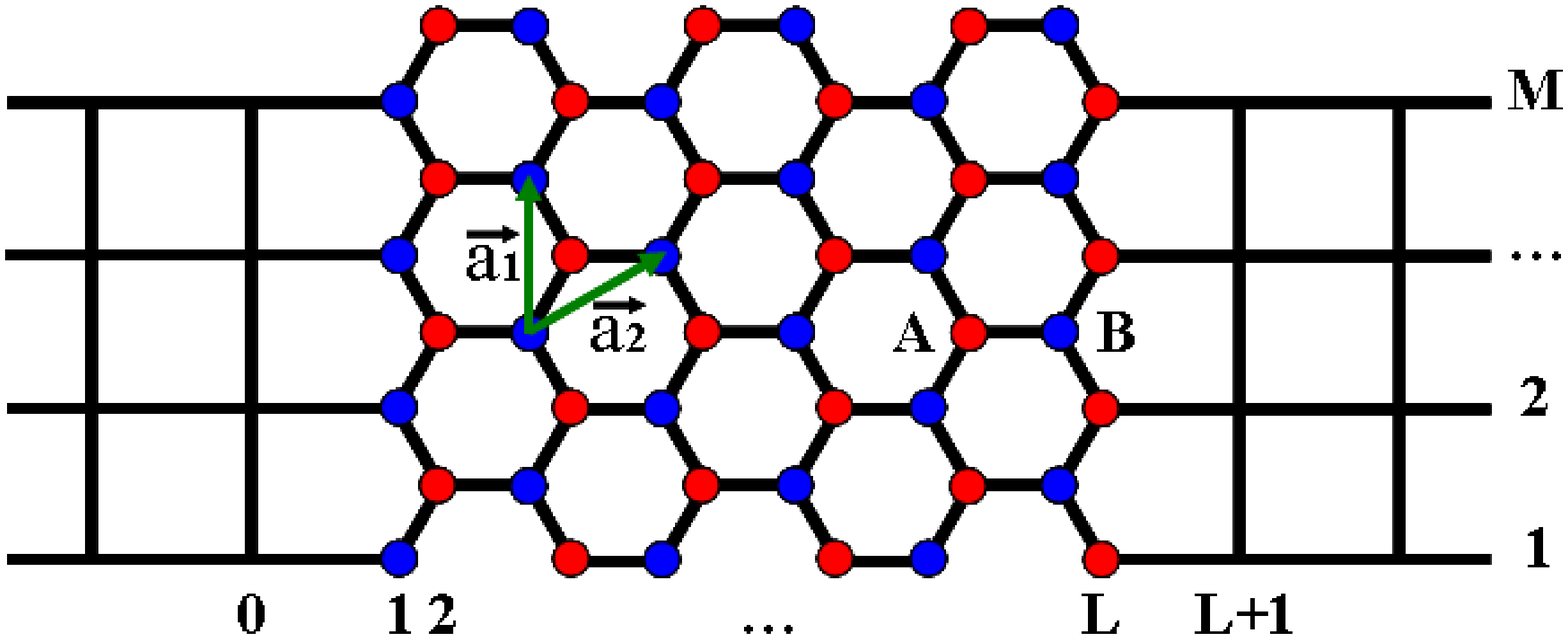}
\caption{\label{fig1} Illustration of armchair graphene nanoribbon (AGR) between two semi-infinite quantum wire contacts. The red and blue circle stand for A and B sublattice in graphene ribbon respectively. Two inequivalent lattice vectors $\vec{a_{1}}$ and $\vec{a_{2}}$ are also shown. The shape of GNR
is determined by $L$ and $M$ carbon atoms in the $x$ and $y$ direction, respectively. Correspondingly, the length is $L_x=\frac{\sqrt{3}La}{4}$ and the width is $L_y=(M-0.5)a$ with the lattice constant $a$ being 2.46\AA.}
\end{figure}

The geometry of AGR sandwiched by two normal metal contacts is illustrated in Fig. 1, where semi-infinite QW stands for the normal contact. The shape of AGR is determined by $L$ and $M$ carbon atoms in the $x$ and $y$ direction, respectively. Correspondingly, the length is $L_x=\frac{\sqrt{3}La}{4}$ and the width is $L_y=(M-0.5)a$, where $a=2.46\AA$ is the lattice constant of graphene. Lattice sites in graphene are classified as two sublattices according to their topological structures with respect to their neighboring sites.

\begin{figure}[tbh]
\includegraphics[width=8.0cm]{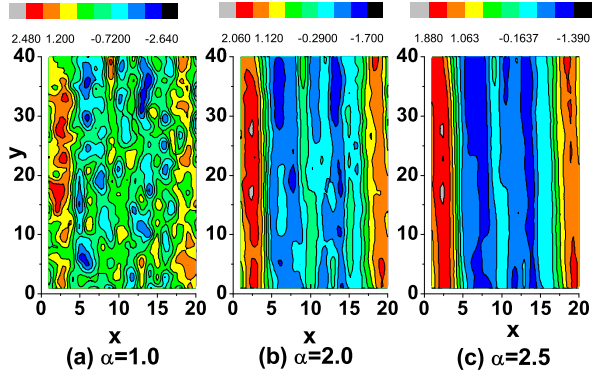}
\caption{\label{fig2} On site energy in the presence of different long-range correlated disorder. The system size is $20 \times 40$. $\alpha$ is 1.0 (a), 2.0 (b) and 2.5 (c). The strength of disorder is $\delta=1$.}
\end{figure}

The transport property of GNR is mainly contributed by $\pi$ orbital and the tight-binding approximation is valid to describe the electronic structure of GNR at the Fermi energy. The Hamilitonian is expressed by
\begin{equation}
H=t\sum_{<ij,i'j'>}C^{\dag}_{ij}C_{i'j'}+\sum_{ij}(\epsilon_{ij}-\mu)C^{\dag}_{ij}C_{ij},
\end{equation}
where $i$ and $j$ are $x$- and $y$-index of lattice site respectively, the nearest neighbors hopping $t$ is set as the energy unit, $<ij,i'j'>$ denotes two nearest neighboring lattice site, $\mu$ is the chemical potential shift by the gate voltage applied to AGR and is zero at QW contacts, and $C^{\dag}_{ij}$ ($C_{ij}$) are electron creation (annihilation) operator at the lattice site, respectively.
$\epsilon_{ij}$ is the onsite energy at lattice site $(i,j)$, which is zero at QW contacts and disordered with long-range correlation at AGR. Two parameters $\alpha$ and $\delta$ measure the strength of correlation and disorder respectively. As shown in Ref. \cite{ZhangGP-EPJB2002}, the onsite energy is defined as follows,
\begin{equation}
\epsilon_{ij}=\sum_{k=1}^{N/2}\left[k^{-\alpha}\left|\frac{2\pi}{N}^{1-\alpha}\right|\right]^{\frac{1}{2}}
\cos\left(\frac{2\pi l_{ij} k}{N}+\phi_{k}\right),
\end{equation}
Here $N=L \times M$ is the total lattice sites of AGR, and $l_{ij}=(i-1)M+j$ is the sequencing of lattice sites that onsite energies at one column of lattice sites are as close as possible. The sequence of onsite energies is normalized with zero mean value $<\epsilon>=0$ and with the variance  $\Delta\epsilon=\sqrt{<\epsilon^2>-<\epsilon>^2}=\delta$.
The disordered onsite energies with long-range correlation are shown in Fig. \ref{fig2}. The disorder strength $\delta$ is set as 1 and the parameter $\alpha$ determines the correlation strength between any two disordered onsite energies. The onsite energies are random as $\alpha$ is 1.0, and become smoother and finally similar to a sequence of stripped potential when $\alpha$ increases from 1.0 to 2.5. As shown in Fig. \ref{fig2}(c), AGR is equivalent to a composite of several random p-n junctions. The hard-wall boundary condition is adopted outside of the system, and the conductance in multichannels' form is calculated as shown in Ref. \cite{YinY-PLA2003}. The Fermi energy $E=0$ is chosen as there are maximal channels at the band center of QW contacts.

\section{Numerical results and discussion}
\subsection{The scaling behavior of the conductance at neutral point in the presence of long-range correlated disorder}
\begin{figure}[tbh]
\includegraphics[width=8.0cm]{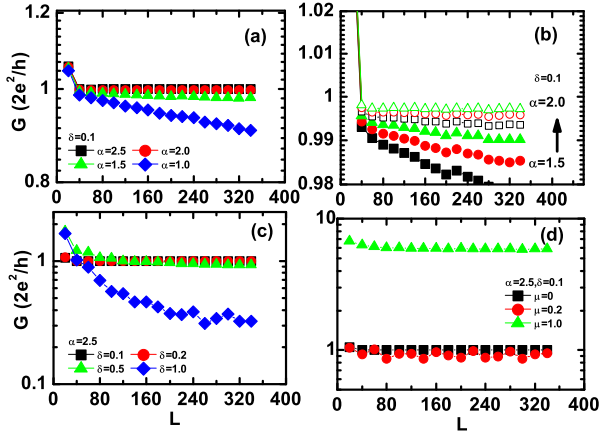}
\caption{\label{fig3} Dependence of the conductance of neutral AGR on the length when $\alpha$ varies at $\delta=0.1$ (a-b) and when $\delta$ varies at $\alpha=2.5$ (c). (d) shows the conductance at $\mu$=0 (square), 0.2 (circle) and 1.0 (upper triangle) with $\alpha=2.5$ and $\delta=0.1$. The width of sample is $10$.}
\end{figure}
It is well known that the band gap of uniform AGR shows a three-family behavior from tight binding model and first principle calculation \cite{SonYW-RPL2006-GNR-TBvasp}, where these two methods predicted that AGR with $N_{a}=3p+2$ as $p$ being an integer has no band gap and finite band gap inversely proportional to the width respectively. Further it was proved that the band gap of AGR with $N_{a}=3p+2$ approaches to zero when the width is wider than 80 $\AA$ through divide-and-conquer approach based on quasi-atomical minimal basis orbits (QUAMBOs) \cite{YaoYX-JPCM2009}. $N_{a}$ is $2M$ in AGR as shown in Fig. \ref{fig1}, and $M=3p+1$ stands for metallic AGR. Even for semiconducting AGRs with $M\ne 3p+1$, the conductance and the localization length always increases with $\alpha$ in the presence of long-range correlated disorder. For sake of simplicity, metallic uniform AGRs are chosen to investigate the effect of long-range correlated disorder here.

Fig. \ref{fig3} shows the dependence of the conductance on the length of AGR in the presence of several different correlated disorder. As shown in Fig. \ref{fig3}(a-b), the conductance approaches to one conductance quantum at highly correlated weak disorder as the length increases, when $\alpha$ is larger than 2.0 at $\delta=0.1$. This proves that long-range correlation in disordered onsite energies incudes a metallic phase in disordered AGR. On the other hand, the conductance exponentially decays as a function of the length like $G \propto \exp(-aL)$. It implies Anderson localization and $a$ is the inverse of the localization length. $a$ is zero for metallic states and finite for localized states depending on the parameters $\alpha$ and $\delta$. In Fig. \ref{fig3}(c), the scaling behavior of the conductance shows that there also exists metal-insulator transition induced by the disorder strength of correlated disorder. From finite size scaling behavior, it is easy to determine the critical $\alpha_{cr}$ in term of $\delta$ and the critical $\delta_{cr}$ in term of $\alpha$ from Fig. \ref{fig3}(a-c) respectively. As shown in Fig. \ref{fig3}(b), $\alpha_{cr}$ is 1.8$\sim$1.9 at $\delta=0.1$. The dependence of conductance on the length of long AGR shows $\alpha_{cr}=1.86$, which is close to but a little larger than the value from the Dirac equation \cite{Cheraghchi-PRB2011}. Moreover, the conductance of Dirac fermion under disordered potential with long-range correlation scales as $G \propto L^{-\eta}$, in which $\eta$ is a constant and depends on $\alpha$ and $\delta$. The discrepancy comes from the difference of tight binding model and Dirac equation, however both tight binding model and Dirac equation predict metal-insulator transition induced by the disorder with long-range correlation. Different from neutral AGR, the conductance oscillates around a fixed value and monotonically decay at the energy shift $\mu$ is 0.2 and 1.0 respectively when the length increases from $20$ to $340$, as shown in (d).

\subsection{Dependence of the conductance on the energy shift in the presence of long-range correlated disorder}
\begin{figure}[tbh]
\includegraphics[width=8.0cm]{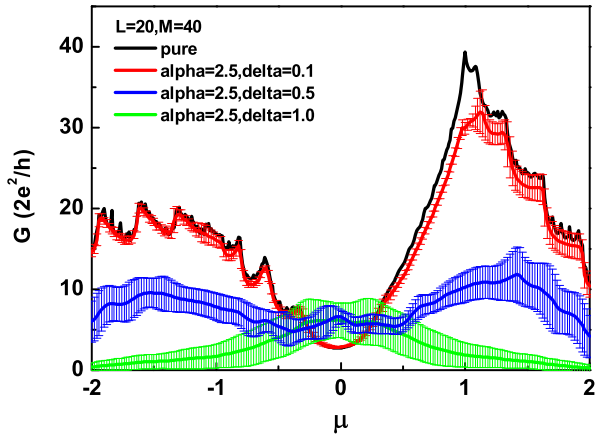}
\caption{\label{fig4} Dependence of the conductance on the energy shift when $\delta$ changes from 0 to 1.0. $\alpha$=2.5. The length and width of sample is $L=20$ and $M=40$ respectively.}
\end{figure}
Fig. \ref{fig4} shows how the conductance varies with the energy shift at three different strengths of disorder. Here energy shift in unit of the nearest-neighbor hopping amplitude $t$ is linear to the gate voltage applied to AGR. The size parameter for AGR is $L=20$ and $M=40$, and $\alpha=2.5$. The conductance of disordered AGR is the average of 500 samples. For comparison, the conductance of uniform AGR is also shown. The slope of conductance at positive energy shift is different from the counterpart at negative energy shift, since the electron-hole symmetry is broken due to odd-numbered ring at the AGR-QW interface \cite{ZhangGP-CPL2011}. For weak disorder, e.g., $\delta=0.1$, the conductance curve is slightly different from that of uniform AGR, and asymmetry in electron-hole still presents. As the disorder strength $\delta$ increases to 0.5, the conductance decreases for most energy shift while increases around zero energy shift. The variance of conductance increases with the disorder strength. When $\delta$ is 1.0, only the conductance near zero energy shift is highest, which is completely different from those above curves of the conductance versus the energy shift. Obviously strong localization occurs as the energy shift is far away from zero at $\delta=1.0$.

\begin{figure}[thb]
\includegraphics[width=8.0cm]{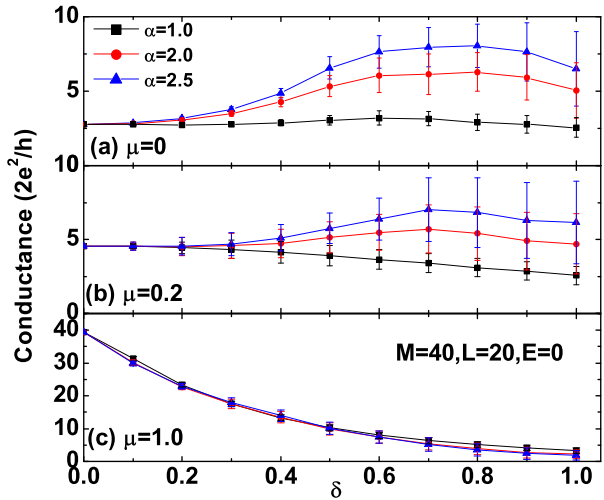}
\caption{\label{fig5} Dependence of the conductance in AGR under several energy shift on the disorder strength $\delta$ when $\alpha$ is 1.0, 2.0 and 2.5 respectively. The length and width of sample is $L=20$ and $M=40$ respectively.}
\end{figure}
Fig. \ref{fig5} shows the dependence of the conductance $G$ at three different energy shifts on the disorder strength $\delta$. For comparison, $\alpha=$1.0, 2.0 and 2.5 are chosen. As shown in Fig. \ref{fig5}(a), the conductance of neutral AGR slightly changes with $\delta$ at $\alpha=1.0$, except that the variance of conductance increases with $\delta$. The conductance first increases and then decreases as $\delta$ increaes when $\alpha$ is larger than 2.0. This behavior is similar to that in the presence of long-range disorder \cite{Rycerz-EPL2007}. On the contrary, the conductance first remains the same at weak disorder and then decay at strong disorder in the presence of short-range disorder such as Anderson disorder. When the energy shift is 0.2, the conductance decreases as $\delta$ increases for lowly correlated disorder, and the behavior of the conductance in the presence of highly correlated disorder is similar to that of neutral AGR , as shown in Fig. \ref{fig5}(b). When the energy shift is 1.0, the conductance decreases as the disorder strength increases for all $\alpha$ as shown in Fig. \ref{fig5}(c). This is due to Anderson localization in heavily dopped AGR.

\begin{figure}[thb]
\includegraphics[width=8.0cm]{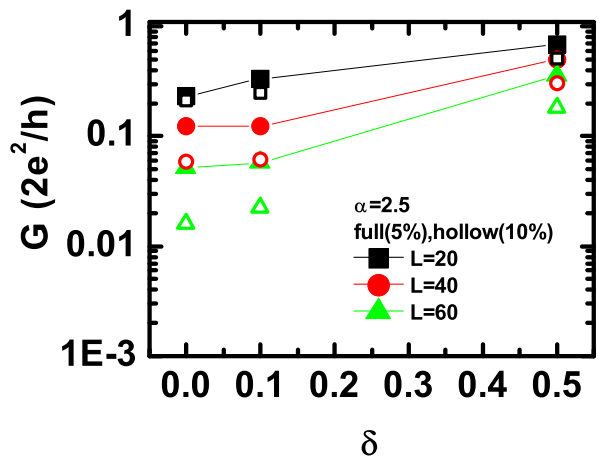}
\caption{\label{fig66} Dependence of the conductance on the disorder strength $\delta$ as the length $L$ varies from 20 to 60. Full and hollow symbols represent 5\% and 10\% edge disorder. $\alpha$=2.5, and the width of sample is $M=4$.}
\end{figure}

It is well known that edge disorder in graphene induces insulating behavior in graphene \cite{edge-disorder}. Different from complex edge disorder, a simple edge disordered graphene nanoribbon is constructed, in which 5\% carbon atoms (i.e. 0.05L and 0.05L) are added randomly to carbon atoms at the top and bottom edges respectively. Then we analyze the conductance of 100 such random samples in the presence of long-range correlated disorder with $\alpha=2.5$. For each random sample, the conductance is averaged between 500 realizations of long-range correlation disorder, as we did for the case without edge disorder. On one hand, the conductance increases with the disorder strength $\delta(\le 0.5)$, as shown in Fig. \ref{fig66}. On the other hand, the conductance decreases as the length increases for all $\delta$, which no delocalization occurs in the presence of highly long-range correlated disorder. This is contrary to the fact that $\alpha=2.0$ induces delocalization in long-range correlated disordered graphene nanoribbon without edge disorder. Higher percent edge disorder results in lower conductance, while the dependence of conductance on $\delta$ is similar. It indicates that no delocalization occurs in edge-disordered graphene nanoribbon even when the disorder is correlated, since localization is induced by edge disorder.

\subsection{Dependence of the conductivity on the energy shift and on the aspect ratio in the presence of long-range correlated disorder}
\begin{figure}[thb]
\includegraphics[width=8.0cm]{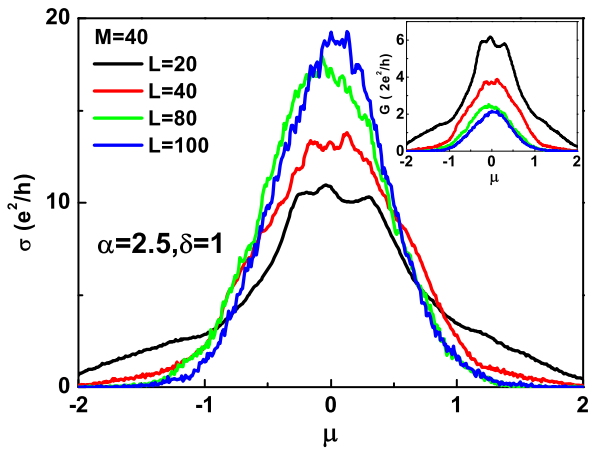}
\caption{\label{fig6} Dependence of the conductivity and the conductance (in the inset) on the energy shift as the length $L$ varies from 20 to 100. $\alpha$=2.5 and $\delta$=1.0. The width of sample is $M=40$.}
\end{figure}
It is well known that the conductivity of uniform and neutral armchair graphene nanoribbon approaches $\frac{4e^{2}}{\pi h}$ as the aspect ratio $M/L$ is far larger than 1 \cite{transport-EPJB2006,transport-PRL2006}. The conductivity is defined as $\sigma=\frac{GL_{x}}{L_{y}}$. Fig. \ref{fig6} shows the dependence of the conductivity on the energy shift in the presence of correlated disorder with $\alpha=2.5$ and $\delta=1.0$. Under such parameters as studied in Sec. III (A), all electronic states are localized at the thermodynamic limit and the conductance decreases as the length increases as shown in the inset of Fig. \ref{fig6}, therefore the conductivity approaches to zero as the aspect ratio approaches to zero. However, for moderate aspect ratio, the conductivity increases at $\left|\mu\right| \le 0.75$ and decreases otherwise as the length increases from 20 to 100, as shown in Fig. \ref{fig6}. The transition of conductivity in term of the length indicates that the conductance scales as $L^{-\beta}$ with $\beta$ varying from less than 1 to larger than 1, because the localization length modulated by the energy shift is longer than and shorter than the length of the system $L$ respectively. We further find that this kind of transition of conductivity occurs in a series of AGRs with different width.

\begin{figure}[thb]
\includegraphics[width=8.0cm]{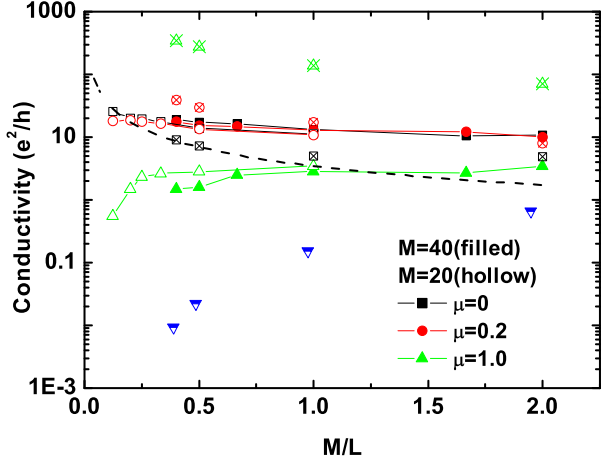}
\caption{\label{fig7} Dependence of the conductivity on the aspect ratio $M/L$. $\alpha$=2.5 and $\delta$=1.0. The energy shift $\mu$ is 0 (square), 0.2 (circle) and 1.0 (upper triangle). The width of sample is $20$ (hollow symbols) and $40$ (filled symbols). Square, circle and upper triangle with cross stand for the conductivity of uniform AGR with $M=40$. Half-filled lower triangle stands for uniform neutral AGR with $M=39$. The dotted black line is the conductivity corresponding to one quantized conductance.}
\end{figure}
Fig. \ref{fig7} shows the dependence of the conductivity on the aspect ratio in the presence of disorder with long-range correlation with $\alpha=2.5$ and $\delta=1.0$. Interestingly, there are two different behaviors for conductivity in term of the aspect ratio. Thereinto, the conductivity near zero energy shift ($\mu$=0 and 0.2) first increases and then decreases while the counterpart at $\mu$=1.0 monotonically decreases, as the aspect ratio approaches to zero. For comparison, the corresponding behavior of the conductivity for uniform AGR with $M=40$ is also shown. Since the conductance remains finite and constant for metallic AGR, the conductivity scales as $O(L/M)$ as $M/L$ approaches to zero for all three different gate voltages. The conductivity corresponding to one quantized conductance is shown by the dotted black line. It is easy to find that the asymptotical conductivity of disordered AGR with long-range correlation is different from metallic uniform AGR, since localization occurs at the thermodynamic limit. On the other hand, the conductivity of disordered AGR at $\mu$=1.0 decreases slower than that of neutral AGR with $M=39$, which is semiconducting without impurity or disorder, while their asymptotical behavior of the conductivity should be the same because of localization. Compared with pure AGR, the conductivity of disordered AGR with long-range correlation depends smoothly on the aspect ratio in general.

\section{Summary}
Transport in disordered armchair graphene nanoribbons (AGR) with long-range correlation between quantum wire contact
is investigated based on tight binding model. Metal-insulator transition is induced by disorder in neutral AGR. Thereinto, the conductance is one conductance quantum for metallic phase and exponentially decays otherwise, when the length approaches to infinity and is far longer than the width. Similar to the case of long-range disorder, the conductance of neutral AGR first increases and then decreases while the conductance of doped AGR monotonically decreases, as the disorder strength increases. In the presence of strong disorder, the conductivity depends monotonically and non-monotonically on the aspect ratio for heavily doped and slightly doped AGR respectively. For edge disordered graphene nanoribbon, the conductance increases with the disorder strength of long-range correlated disordered while no delocalization occurs, since localization is induced by the edge disorder.

\section*{Acknowledgments}
This work was supported by the Research Foundation from Ministry of Education of China (2009EDU309002), the National Basic Research Program of China (2012CB921704), and the National Natural Science Foundation of China under Grant No 11174363.

\end{document}